\documentclass[a4paper,12pt]{amsart}
\usepackage[utf8]{inputenc}
\usepackage[all]{xy}
\usepackage{amssymb}
\usepackage{amsmath}
\usepackage{csquotes}
\usepackage{subfig}
\usepackage{enumerate}
\usepackage{graphicx}
\usepackage{enumerate}
\usepackage{amsaddr}

\usepackage{color}
\usepackage{setspace}   

\usepackage[left=1in, right=1in]{geometry}
\usepackage{subfig}

\usepackage{xcolor}

\usepackage{soul}

\usepackage{cancel}

\usepackage{setspace}
\title{On null hypotheses in survival analysis.}
\author{Mats J. Stensrud, Kjetil Røysland and Pål C. Ryalen}
\address{Department of Biostatistics, University of Oslo, Domus Medica Gaustad, Sognsvannsveien 9, 0372 Oslo, Norway}
\date{\today}

\MakeOuterQuote{"}\EnableQuotes
\DeclareUnicodeCharacter{00A0}{ }

\begin{document}
\textwidth=397.485pt

\begin{abstract}
The conventional nonparametric tests in survival analysis, such as the log-rank
test, assess the null hypothesis that the hazards are equal at all times.
However, hazards are hard to interpret causally, and other null hypotheses are more relevant in many scenarios with
survival outcomes. To allow for a wider range of null hypotheses, we present a
generic approach to define test statistics. This approach utilizes the fact that a wide range of common parameters in survival analysis can be expressed as solutions of differential equations. Thereby we can test 
hypotheses based on survival parameters that solve differential equations driven by cumulative hazards,
and it is easy to implement the tests on a computer. We present simulations, suggesting that our tests perform well for
several hypotheses in a range of scenarios. Finally, we use our tests to evaluate the effect of adjuvant chemotherapies in patients with colon cancer, using data from a randomised controlled trial.




\end{abstract}
\maketitle

\textit{KEY WORDS: } Causal inference; Hazards; Hypothesis testing; Failure time analysis. 

\clearpage

\section{Introduction}


The notion of hazards has been crucial for the development of modern
survival analysis.  Hazards are perhaps the most natural parameters to use
when fitting statistical models to time-to-event data subject to censoring,
and hazard functions were essential for the development of popular methods like
Cox regression and rank tests, which are routinely used in practice.

In the field of causal inference, however, there is concern that many 
statisticians just do advanced 'curve fitting' without being careful about the interpretation of the parameters that are reported \cite{pearl1997new,hernanRobins, pearl}. 
This criticism can be directed to several areas in statistics. In this spirit, we think that statisticians in general should pay particular attention to effect measures with clear-cut causal interpretations. 
        
In survival analysis, it has been acknowledged that interpreting hazards as effect measures is	delicate, see e.g.\ \cite{aalen2015does} and
\cite{hernan2010hazards}. This contrasts the more traditional opinion, in which the proportional hazards model is motivated by the 'simple and easily understood interpretation' of hazard ratios \cite[4.3.a]{coxOakes}.
A key issue arises because the hazard, by definition, is conditioned on previous
survival.  If we consider causal diagrams \cite{pearl,hernanRobins}, it is clear that we condition on a 'collider' that opens a non-causal pathway from the
exposure through any unobserved heterogeneity into the event of interest, see
\cite{aalen2015does,stensrud2017exploring, ryalen2018additive}.  Since unobserved heterogeneity is present in most practical scenarios, even in randomized
trials, the conditioning means that the hazards are fundamentally hard to interpret causally \cite{aalen2015does, hernan2010hazards}. 

Although we must be careful about assigning causal interpretations to hazards, we do not claim that hazards are worthless. On the contrary, hazards are key elements in the modelling of other parameters that are easier to interpret, serving as building blocks. This point of view is also found in
\cite[17.1]{hernanRobins}:
        "..., the survival analyses in this book privilege
        survival/risk over hazard. However, that does not mean that we should
        ignore hazards. The estimation of hazards is often a useful
        intermediate step for the estimation of survivals and risks." Indeed, we have recently suggested a generic method to estimate a range of
effect measures in survival analysis, utilizing differential equations driven by cumulative hazards
\cite{ryalen2017transforming}. 

Nevertheless, the conventional hypothesis tests in survival analysis are still based on hazards. In particular the rank tests \cite{aalen1978nonparametric}, including the log-rank test, are based on the null hypothesis
\begin{align} 
\textbf{H}_0 \text{: } \alpha_t^{1} = \alpha_t^{2} \text{ for all } t \in [0, \mathcal T],
\label{eq: H0 log-rank}
\end{align}
where $\alpha_t^i$ is the hazard in group $i$. 
Formulating such hypotheses in a practical setting will often imply that we
assign causal interpretations to these hazard functions. In the simplest survival setting this is not a
problem, as there is a one-to-one relationship between hazards and the
survival curves, and a null hypothesis comparing two or more survival curves is
straightforward. In more advanced settings, e.g.\ scenarios with competing risks,
hypotheses like \eqref{eq: H0 log-rank} are less transparent, leading to issues with interpretation \cite{young2018choice}. For example, in competing risks settings where competing events are treated as censoring events, the null hypothesis in \eqref{eq: H0 log-rank} is based on cause-specific hazards, which are often not the target of inference \cite{young2018choice}. 

We aimed to develop new hypothesis tests for time-to-event outcomes with two key characteristics: First, the tests should be rooted in explicit null hypotheses that are easy to interpret. Second, the testing strategy should be generic, such that the scientist can apply the test to their estimand of interest.

\section*{Survival parameters as solutions of differential equations}
\label{section: param as solutions to diff.eq}

We will consider survival parameters that are functions solving differential equations on the form 
\begin{align}
	X_t &= X_0 + \int_0^t F(X_s) dA_s, \label{eq:ode}
\end{align}
where $A$ is a $q$ dimensional vector of cumulative hazards, and $F = (F_1,\cdots,F_q): \mathbb{R}^p \longrightarrow \mathbb{R}^{p \times q}$ is Lipschitz continuous with bounded and continuous first and second derivatives, and satisfies a linear growth bound. The class of parameters also includes several quantities that are Lebesgue integrals, such that $dA^i_t = dt$ for some $i$. Here, $X$ is a vector that includes our estimand of interest, but $X$ may also contain additional nuisance parameters that are needed to formulate the estimand of interest.


Many parameters in survival analysis solve equations on the form \eqref{eq:ode}. In particular, the survival function can be expressed on the form \eqref{eq:ode} as $S_t = 1 - \int_0^t S_s dA_s$, where $A$ is the cumulative hazard for death. Other examples include the cumulative incidence function, the restricted mean survival function, and the prevalence function. We will present these parameters in detail in Section \ref{sec: test statistic examples}. Nonparametric plugin estimators have been thoroughly studied in \cite{ryalen2017transforming}. The strategy assumes that $A$ can be consistently estimated by
\begin{align}
	\hat A_t &= \int_0^t G_{s-} dN_s, \label{eq:cumulative hazard pure jump estimator}
\end{align}
where $G$ is a $q \times l$ dimensional predictable process, and $N$ is an $l$ dimensional counting process. Furthermore, we assume that $\hat A$, the residuals $W^n = \sqrt{n}(\hat A - A)$, and its quadratic variation $[W^n]$, are so-called \emph{predictably uniformly tight}. When the estimator is a counting process integral, a relatively simple condition ensures predictable uniformly tightness \cite[Lemma 1]{ryalen2017transforming}. Moreover, we suppose that $\sqrt{n}(\hat A - A)$ converges weakly (wrt the Skorohod metric) to a mean zero Gaussian martingale with independent increments, see \cite[Lemma 1, Theorem 1 \& 2]{ryalen2017transforming} for details. Examples of estimators on the form \eqref{eq:cumulative hazard pure jump estimator} that satisfy these criteria are the Nelson-Aalen estimator, or more generally Aalen's additive hazard estimator; if Aalen's additive hazards model is a correct model for the hazard $A$, then Aalen's additive hazard model satisfy these criteria, in particular predictable uniformly tightness.

Our suggested plugin estimator of $X$ is obtained by replacing $A$ with $\hat A$, giving estimators that solve the system
\begin{align}
	\hat X_t &= \hat X_0 + \int_0^t F(\hat X_{s-}) d\hat A_s, \label{eq:sde}
\end{align}
where $\hat X_0$ is a consistent estimator of $X_0$. When the estimand is the survival function, this plugin estimator reduces to the Kaplan-Meier estimator. Ryalen et al \cite{ryalen2017transforming} identified the asymptotic distribution of $\sqrt{n}(\hat X_t - X_t)$ to be a mean zero Gaussian martingale with covariance $V$ solving a linear differential equation \cite[eq. (17) ]{ryalen2017transforming}. The covariance $V$ can also be consistently estimated by inserting the estimates $\hat A$, giving rise to the system
\begin{equation}
        \begin{split}
                \hat V_t  =& \hat V_0  + \sum_{j = 1}^q \int_0^t
                {\hat V}_{s-}  \nabla F_j( {\hat X}_{s-} )^\intercal 
                + \nabla F_j(  {\hat X}_{s-} )   {\hat V}_{s-}  d
                {\hat A}_{s}^{j} \\ &
                + n \int_0 ^t F( { \hat X}_{s-} ) d [B]_s F(  {\hat X}_{s-} )^\intercal,
                        \end{split}
                        \label{eq:XVarEst}
        \end{equation}
where $\{ \nabla F_j \}_{j=1}^q$ are the Jacobian matrices of the columns of $F = (F_1,\cdots,F_q)$ from \eqref{eq:ode}, and $[B]_t$ is a $q \times q$ matrix defined by 
\begin{align*}
    \Big([B]_t\Big)_{i,j}=\begin{cases}
        0, \text{ if } dA_t^i = dt \text{ or } dA_t^j = dt \\
         \sum\limits_{s \leq t} \Delta \hat A_s^i \Delta \hat A_s^j, \text{ otherwise.} 
    \end{cases}
\end{align*}
The variance estimator \eqref{eq:XVarEst}, as well as the parameter estimator \eqref{eq:sde}, can be expressed as difference equations, and therefore they are easy to calculate generically in computer programs. 
To be explicit, let $\tau_1, \tau_2,  \dots $ denote the ordered jump times of $ {\hat A}$. Then, ${\hat X}_t =  {\hat X}_{\tau_{k-1}} + F({\hat X}_{\tau_{k-1}} )  \Delta {\hat A}_{\tau_{k}},$ as long as $\tau_{k} \leq t < \tau_{k+1}$. Similarly, the plugin variance equation may be written as a difference equation,
\begin{align*}
    \begin{split}
             {\hat V}_t = & {\hat V}_{\tau_{k-1}}  + \sum_{j = 1}^q  
            {\hat V}_{_{\tau_{k-1}}}  \nabla F_j( {\hat X}_{\tau_{k-1}} )^\intercal 
            + \nabla F_j(  {\hat X}_{\tau_{k-1}} )   {\hat V}_{\tau_{k-1}}  \Delta
            {\hat A}_{\tau_{k}}^{j} \\ &
            + n F( { \hat X}_{\tau_{k-1}} ) \Delta [   B 
                    ]_{\tau_{k}} F(  {\hat X}_{\tau_{k-1}} )^\intercal.
                    \end{split}
\end{align*}


\section{Hypothesis testing}

The null hypothesis is not explicitly expressed in many research reports. On the contrary, the null hypothesis is often stated informally, e.g.\ vaguely indicating that a difference between two groups is assessed. Even if the null hypothesis is perfectly clear to a statistician, this is a problem: the applied scientist, who frames the research question based on subject-matter knowledge, may not have the formal understanding of the null hypothesis. 

In particular, we are not convinced that scientists faced with time-to-event outcomes profoundly understand how null hypotheses based on hazard functions. Hence, using null hypotheses based on hazard functions, such as \eqref{eq: H0 log-rank}, may be elusive: in many scenarios, the scientist's primary interest is not to assess whether the hazard functions are equal \textit{at all follow-up times}. Indeed, the research question is often more focused, and the scientist's main concern can be contrasts of other survival parameters at a prespecified $t$ or in a prespecified time interval \cite{klein2007analyzing}. For example, we may aim to assess whether cancer survival differs between two treatment regimens five years after diagnosis. Or we may aim to assess whether a drug increases the average time to relapse in subjects with a recurring disease. We will highlight that the rank tests are often not suitable for such hypotheses.

Hence, instead of assessing hazards, let us study tests of (survival) parameters $X^1$ and $X^2$ in groups 1 and 2 at a prespecified time $t_0$. 
The null hypothesis is
\begin{align}
\textbf{H}^{X}_0 \text{: } X_{t_0}^{1,i} = X_{t_0}^{2,i} \text{ for } i=1,\cdots, p,
\label{eq: H0 prespecified t}
\end{align}
where $p$ is the dimension of $X$. We emphasize that the null hypothesis in \eqref{eq: H0 prespecified t} is different from the null hypothesis in \eqref{eq: H0 log-rank}, as \eqref{eq: H0 prespecified t} is defined for any parameter $X_{t_0}$ at a $t_0$. We will consider parameters $X^1$ and $X^2$ that solve \eqref{eq:ode}; this is a broad class of important parameters, including (but not limited to) the survival function, the cumulative incidence function, the time dependent sensitivity and specificity functions, and the restricted mean survival function \cite{ryalen2017transforming}.


\subsection{Test statistics}
\label{sec: test statistics}
We consider two groups 1 and 2 with population sizes $n_1,n_2$ and let $n = n_1 + n_2$. We can estimate parameters $X^1, X^2$ and covariance matrices $V^1, V^2$ using the plugin method described in Section \ref{section: param as solutions to diff.eq}. The contrast $\sqrt{n}(\hat X_{t_0}^1 - \hat X_{t_0}^2)$ has an asymptotic mean zero normal distribution under the null hypothesis. 
If the groups are independent, we may then use the statistic
\begin{align}
	(\hat{X}_{t_0}^1 - \hat{X}_{t_0}^2) \hat{V}_{t_0}^{-1} (\hat{X}_{t_0}^1 - \hat{X}_{t_0}^2)^\intercal, \label{eq: prespecified t statistic independence}
\end{align}
to test for differences at $t_0$, where $\hat{V}_{t_0} = \hat V_{t_0}^1/n_1 + \hat V_{t_0}^2/n_2$, and where $\hat V_{t_0}^1$ and $\hat V^2$ are calculated using the covariance matrix estimator \eqref{eq:XVarEst}. Then, the quantity \eqref{eq: prespecified t statistic independence} is asymptotically $\chi^2$ distributed with $p$ degrees of freedom under the null hypothesis, which is a corollary of the results in \cite{ryalen2017transforming}, as we know from \cite[Theorem 2]{ryalen2017transforming} that $\sqrt{n_i}(\hat X^i - X^i$) converges weakly to mean zero Gaussian martingale whose covariance matrix $V^i$ can be consistently estimated using \eqref{eq:XVarEst}. Therefore, under the null hypothesis \eqref{eq: H0 prespecified t}, the root $n$ difference of the estimates, $\sqrt{n} (\hat X^1_{t_0} - \hat X^2_{t_0})$, will converge to a multivariate mean zero normal distribution with covariance matrix that can be estimated by $n \big( \hat V^1_{t_0} /n_1 + \hat V^2_{t_0} / n_2 \big)$. Due to the continuous mapping theorem, the statistic \eqref{eq: prespecified t statistic independence} has an asymptotic $\chi^2$ distribution.



Sometimes we may be interested in testing e.g.\ the $r$ first components of $X^1$ and $X^2$, under the null hypothesis $X_{t_0}^{1,i} = X_{t_0}^{2,i}$ for $i=1,\cdots , r < p$. It is straightforward to adjust the hypothesis \eqref{eq: H0 prespecified t} and the test statistic, yielding the same asymptotic distribution with $r$ degrees of freedom.

\section{Examples of test statistics}
\label{sec: test statistic examples}
We derive test statistics for some common effect measures in survival and event history analysis. By expressing the test statistics explicitly, our tests may be compared with the tests based on conventional approaches. 




\subsection{Survival at $t_0$}
\label{sec: estim surv}
In clinical trials, the primary outcome may be survival at a prespecified $t$, e.g.\ cancer survival 5 years after diagnosis. Testing if survival at $t$ is equal in two independent groups can be done in several ways \cite{klein2007analyzing}, e.g.\ by estimating the variance of Kaplan-Meier curves using Greenwood's formula. However, we will highlight that our generic tests also immediately deal with this scenario:\ it is straightforward to use the null hypothesis in \eqref{eq: H0 prespecified t}, where $S_t^{1}$ and $S_t^{2}$ are the survival functions in group $1$ and $2$ at time $t$. Using the results in Section \ref{sec: test statistics}, we find that the plugin estimators of $S^1$ and $S^2$ are the standard Kaplan-Meier estimators. 
The plugin variance in group $i$ solves
\begin{align}
\label{eq: survival plugin variance}
    \hat{V}_t^i = \hat{V}^{i}_{0} - 2 \int_0^t\hat{V}^{i}_{s-} d\hat{A}^i_{s} + n_i \int_0^t \Big( \frac{\hat{S}^i_{s-}}{Y_s^i} \Big)^2 d N_s^i,
\end{align}
for $i \in \{1,2\}$, where $Y_s^i$ is the number at risk in group $i$ just before time $s$. 
Assuming that the groups are independent, the final variance estimator can be expressed as $\hat {V}_t =  \hat{V}_t^1/n_1 + \hat{V}_t^2/n_2$, and the statistic \eqref{eq: prespecified t statistic independence} becomes $	(\hat S^1_{t_0} - \hat S^2_{t_0})^2 / \hat{V}_{t_0} $, which is approximately $\chi^2$ distributed with 1 degree of freedom.



\subsection{Restricted mean survival until $t_0$}
\label{sec: estim rmst}
As an alternative to the hazard ratio, the restricted mean survival has been advocated: it can be calculated without parametric assumptions and it has a clear causal interpretation \cite{royston2011use,trinquart2016comparison,tian2017efficiency}. The plugin estimator of the restricted mean survival difference between groups $1$ and $2$ is
    $\hat R_t^1 - \hat R_t^2 = \sum_{\tau_k \leq t} \big( \hat S_{\tau_{k-1}}^1 - \hat S_{\tau_{k-1}}^2 \big) \Delta \tau_k,$
where $\Delta \tau_k = \tau_k - \tau_{k-1}$. The plugin estimator for the variance is
\begin{align*}
	\hat V_t^{R^i} &= \hat V_0^{R^i} + 2 \sum_{\tau_k \leq t} \hat V_{ \tau_{k-1}}^{R^i,S^i} \Delta \tau_k \\
    \hat V_t^{R^i,S^i} &= \hat V_0^{R^i,S^i} - \int_0^t \hat V_{s-}^{R^i,S^i} d \hat A_s^i + \sum_{\tau_{k} \leq t}  \hat V^{S^i}_{\tau_{k-1}} \Delta \tau_k,
\end{align*}
where $\hat V^{S^i}$ is the plugin variance for $\sqrt{n_i}\hat{S}^i$, given in \eqref{eq: survival plugin variance}. The statistic \eqref{eq: prespecified t statistic independence} can be used to perform a test, with $\hat V_{t_0} = \hat V^{R^1}_{t_0}/n_1 + \hat V^{R^2}_{t_0}/n_2$. 



\subsection{Cumulative incidence at $t_0$}
\label{sec: cum inc}
Many time-to-event outcomes are subject to competing risks. The Gray test is a competing risk analogue to the log-rank test: the null hypothesis is defined by subdistribution hazards $\lambda_t$, such that $\lambda_t = \frac{d}{dt}  \log [1-C_t]$ where $C_t$ is the cumulative incidence of the event of interest, are equal at all $t$ \cite{gray1988class}. Analogous to the log-rank test, the Gray test has low power if the subdistribution hazard curves are crossing \cite{latouche2007sample}. However, we are often interested in evaluating the cumulative incidence at a time $t_0$, without making assumptions about the subdistribution hazards, which are even harder to interpret causally than standard hazard functions. 
By expression the cumulative incidence on the form \eqref{eq:ode}, we use our transformation procedure to obtain a test statistic for the cumulative incidence at $t_0$. The plugin estimator for the cumulative incidence difference is
\begin{align*}
	\hat C_{t}^1 -\hat C_{t}^2 &= 
    \int_0^t \hat S_{s-}^1 d\hat A^{1,j}_s - \int_0^t \hat S_{s-}^2 d\hat A^{2,j}_s,
\end{align*}
where $A^{i,j}$ is the cumulative cause-specific hazard for the event $j$ of interest, and $\hat S^i$ is the Kaplan-Meier estimate within group $i$.
The groupwise plugin variances solve
\begin{align*}
	\hat V^i_t &= \hat V^i_{0} + 2 \int_0^t \hat V_{s-}^i d \hat A^{i,j}_{s} + n_i \int_0^t  \Big( \frac{\hat S^i_{s-}}{Y_s^i} \Big)^2 d N_s^{i,j},
\end{align*}
where $N^{i,j}$ counts the event of interest. 

\subsection{Frequency of recurrent events} 
\label{sec: freq rec ev}
Many time-to-event outcomes are recurrent events. For example, time to hospitalization is a common outcome in medical studies, such as trials on cardiovascular disease. Often recurrent events are analysed with conventional methods, in particular the Cox model, restricting the analysis to only include the first event in each subject. A better solution may be to study the mean frequency function, i.e.\ the marginal expected number of events until time $t$, acknowledging that the subject can not experience events after death \cite{ghosh2000nonparametric}. 
We let $A^{i,E}$ and $A^{i,D}$ be the cumulative hazards for the recurrent event and death in group $i$, respectively, and let $K^i$ and $S^i$ be the mean frequency function and survival, respectively. Then, the plugin estimator of the difference is
\begin{align*}
	\hat K_t^1 - \hat K_t^2 &= \int_0^t \hat S_{s-}^{1} d\hat A_s^{1,E} - \int_0^t \hat S_{s-}^{2} d\hat A_s^{2,E}.
\end{align*}
The plugin variances solve
\begin{align*}
	\hat V^{K^i}_t &= \hat V^{K^i}_{0}  + \int_0^t\hat V^{K^i,S^i}_{s-} d( \hat A^{i,E}_{s} - \hat A^{i,D}_{s}) + n_i \int_0^t  \Big( \frac{\hat{S}_{s -}^i}{Y^i_s} \Big)^2 d N_s^{i,E}, \\
    \hat V^{K^i,S^i}_t &= \hat V^{K^i,S^i}_{0} - \int_0^t\hat V^{K^i,S^i}_{s-} d \hat A_{s}^{i,D}  + \int_0^t \hat V_{s-}^{S^i} d \hat A^{i,E}_{s},
\end{align*}
where $N^{i,E}$ counts the recurrent event, and where $\hat V^{S^i}$ is the survival plugin variance in group $i$, as displayed in \eqref{eq: survival plugin variance}. 

\subsection{Prevalence in an illness-death model}
\label{sec: prev ill death}
\label{section:prevalence}
The prevalence denotes the number of individuals with a condition at a specific time, which is e.g.\ useful for evaluating the burden of a disease. We consider a simple Markovian illness-death model with three states: healthy:0, ill:1, dead:2. The population is assumed to be healthy initially, but individuals may get ill or die as time goes on. We aim to study the prevalence $P_t^{i,1}$ of the illness in group $i$ as a function of time. Here, we assume that the illness is irreversible, but we could extend this to a scenario in which recovery from the illness is possible, similar to Bluhmki \cite{bluhmki2018wild}.  Let $A^{i,kj}$ be the cumulative hazard for transitioning from state $k$ to $j$ in group $i$. Then, $P^{i,1}$ solves the system
\begin{align*}
	\begin{pmatrix}
		P^{i,0}_t \\ P_t^{i,1}
	\end{pmatrix} &=
    \begin{pmatrix}
    	1\\0
    \end{pmatrix}
    + \int_0^t \begin{pmatrix}
    	-P_s^{i,0} & -P_s^{i,0} & 0 \\
        P_s^{i,0} & 0 & -P_s^{i,1}
    \end{pmatrix}
    d\begin{pmatrix}
    	A^{i,01}_s\\ A^{i,02}_s \\ A^{i,12}_s
    \end{pmatrix}.
\end{align*}
The plugin estimator for the difference $P^{1,1} - P^{2,1}$ is
\begin{align*}
	\hat P^{1,1}_t - \hat P^{2,1}_t &=  \int_0^t \hat P^{1,0}_{s-} d\hat A_s^{1,01} - \int_0^t \hat P^{2,0}_{s-} d\hat A_s^{2,01} - \int_0^t \hat P^{1,1}_{s-} d\hat A_s^{1,01} + \int_0^t \hat P^{2,1}_{s-} d\hat A_s^{2,01}.
\end{align*}
The variance estimator for group $i$ reads
\begin{align*}
	\hat V_t^{P^{i,1}} &= \hat V_{0}^{P^{i,1}} + 2  \int_0^t \hat V_{s-}^{P^{i,0},P^{i,1}} d \hat A_s^{i,01} - 2  \int_0^t\hat V_{s-}^{P^{i,1}} d \hat A_{s}^{i,12} \\
    &+ n_i  \Big( \int_0^t  \Big( \frac{\hat P^{i,0}_{s-}}{Y_s^{i,0}} \Big)^2 dN_s^{i,01} + \int_0^t  \Big( \frac{\hat P^{i,1}_{s-}}{Y_s^{i,1}} \Big)^2 dN_s^{i,12} \Big) \\
    \hat V_t^{P^{i,0},P^{i,1}} &= \hat V_{0}^{P^{i,0},P^{i,1}} + \int_0^t \big( \hat V_{s-}^{P^{i,0}} - \hat V_{s-}^{P^{i,0},P^{i,1}} \big) d \hat A_{s}^{i,01} -  \int_0^t \hat V_{s-}^{P^{i,0},P^{i,1}} d \hat A_{s}^{i,02}  \\
    &- \int_0^t \hat V_{s-}^{P^{i,0},P^{i,1}} d \hat A_{s}^{i,12} - n_i \int_0^t \Big( \frac{\hat P^{i,0}_{s-}}{Y_s^{i,0}} \Big)^2 dN_s^{i,01} \\
    \hat V_t^{P^{i,0}} &= \hat V_{0}^{P^{i,0}} - 2 \int_0^t \hat V_{s-}^{P^{i,0}} d \hat A_{s}^{i,01}- 2 \int_0^t \hat V_{s-}^{P^{i,0}} d \hat A_{s}^{i,02}
    \\ & + n_i \int_0^t \Big( \frac{\hat P_{s-}^{i,0}}{Y_s^{i,0}} \Big)^2 d\big(  N_s^{i,01} + N_s^{i,02} \big).
\end{align*}
Here, $Y^{i,0}, Y^{i,1}$ are the number of individuals at risk in states $0$ and $1$, while $N^{i,kj}$ counts the transitions from state $k$ to $j$ in group $i$. By calculating $\hat V_t^{P^{i,1}}$ for $i \in \{ 1, 2 \}$, we can find the statistic \eqref{eq: prespecified t statistic independence}. Here, the prevalence is measured as the proportion of affected individuals relative to the population at $t=0$. We could use a similar approach to consider the proportion of affected individuals relative to the surviving population at $t$, or we could record the cumulative prevalence until $t$ to evaluate the cumulative disease burden.

\section{Performance}
\label{sec: performance}
In this section, we present power functions under several scenarios for the test statistics that were presented in Section \ref{sec: test statistic examples}. The scenarios were simulated by defining different relations between the hazard functions in the two exposure groups: (i) constant hazards in both groups, (ii) hazards that were linearly crossing, and (iii) hazards that were equal initially before diverging after a time $t$.

For each hazard relation (i)-(iii), we defined several $\kappa$'s such that the true parameter difference was equal to $\kappa$ at the prespecified time point $t_0$, i.e. $X_{t_0}^1 - X_{t_0}^2 = \kappa$. For each combination of target parameter, difference $\kappa$, and hazard scenario, we replicated the  simulations $m$ times to obtain $m$ realizations of \eqref{eq: prespecified t statistic independence}, and we artificially censored 10\% of the subjects in each simulation. In the Supplementary material, we show additional simulations with different sample sizes (50, 100 and 500) and fractions of censoring (10\%-40\%). We have provided an overview of the simulations in figure \ref{fig:simulation scenarios}, in which parameters of interest (solid lines) and hazard functions (dashed lines) are displayed in scenarios with fixed $\kappa=-0.05$ at $t_0 = 1.5$, i.e. $X^1_{1.5} -X^2_{1.5} = -0.05$.

\begin{figure}
\centering
\setlength{\lineskip}{1ex}
\subfloat{\includegraphics[width=.45\textwidth]{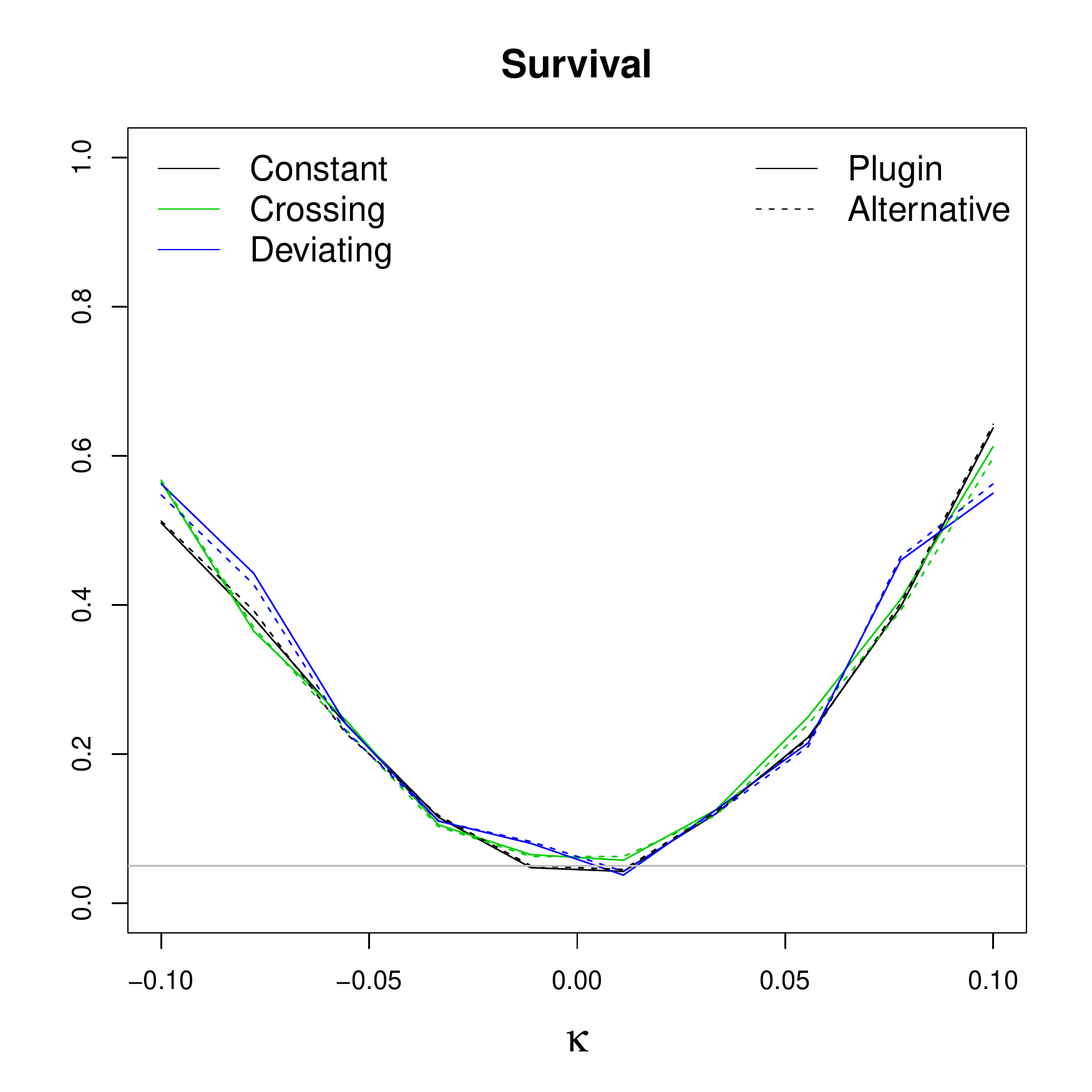}}%
\subfloat{\includegraphics[width=.45\textwidth]{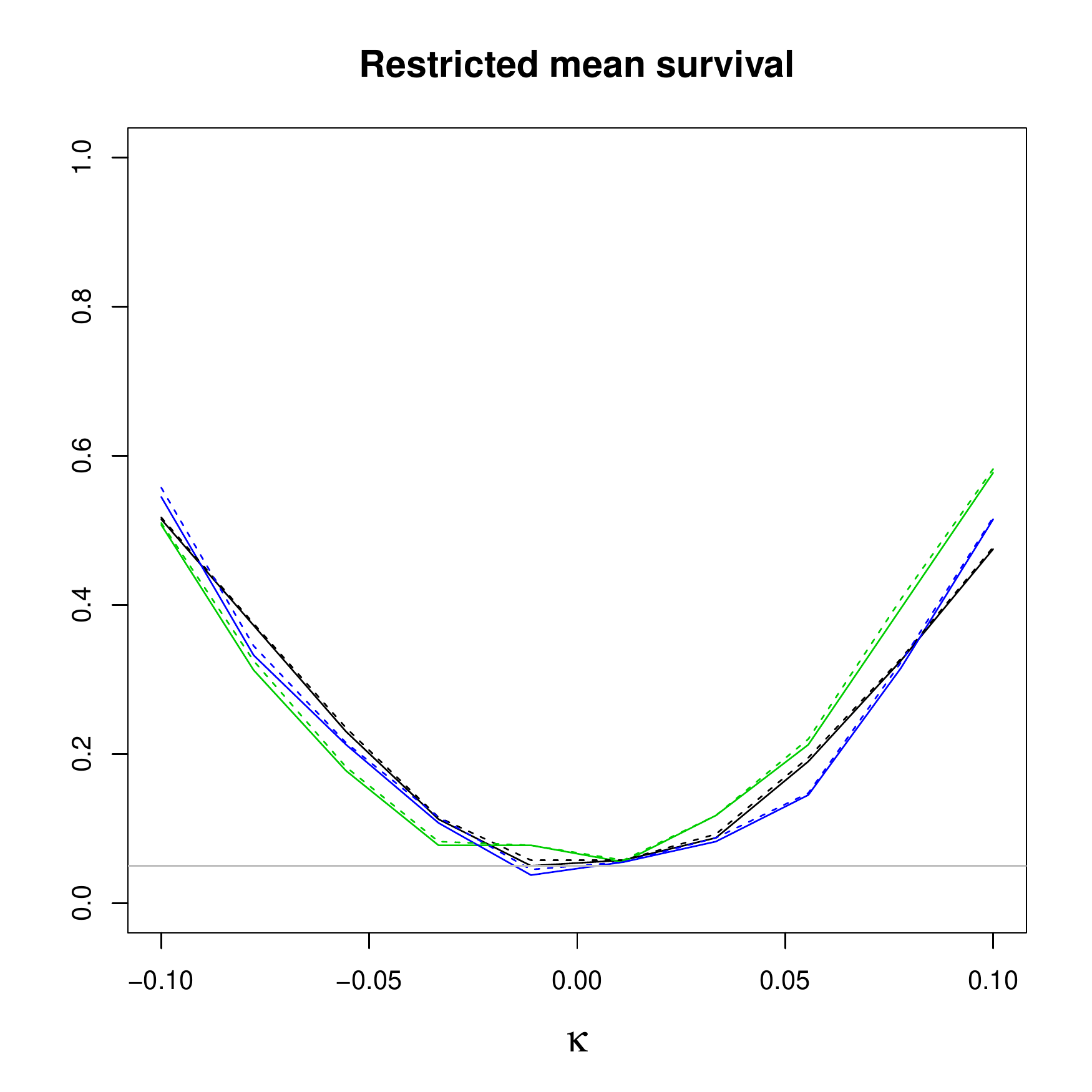}}%
\hspace{.01\textwidth}%
\subfloat{\includegraphics[width=.45\textwidth]{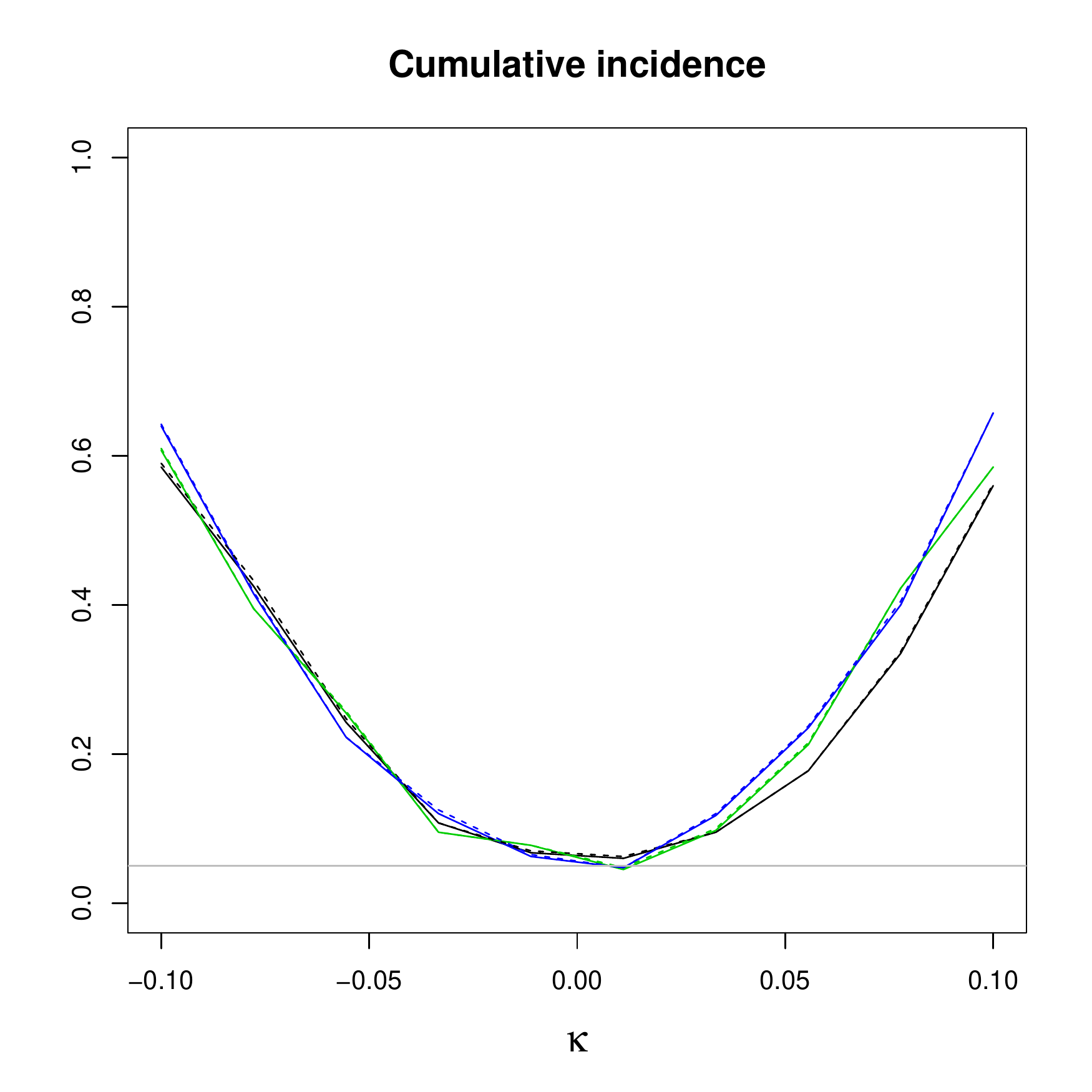}}%
\subfloat{\includegraphics[width=.45\textwidth]{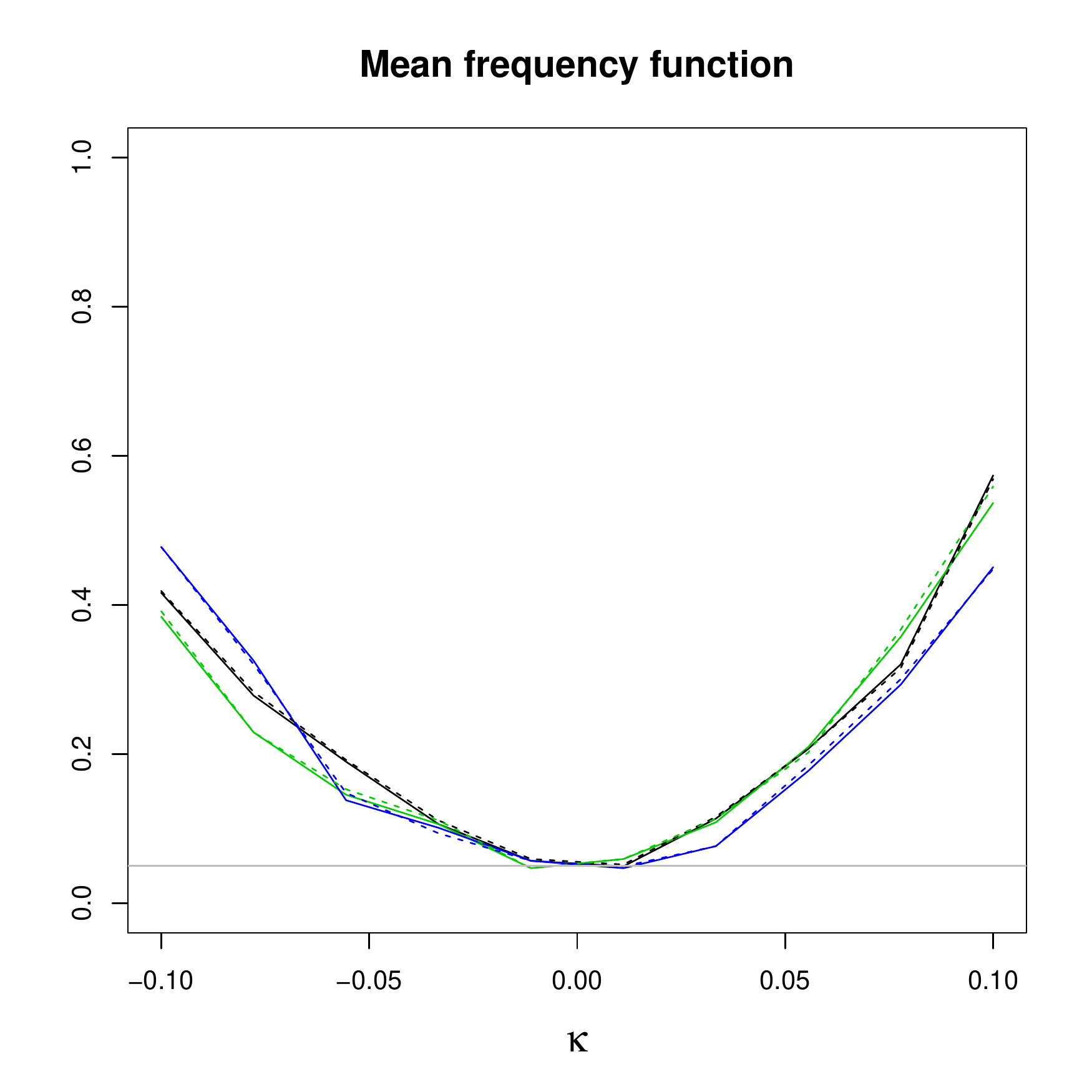}}%
\hspace{.01\textwidth}%
\subfloat{\includegraphics[width=.45\textwidth]{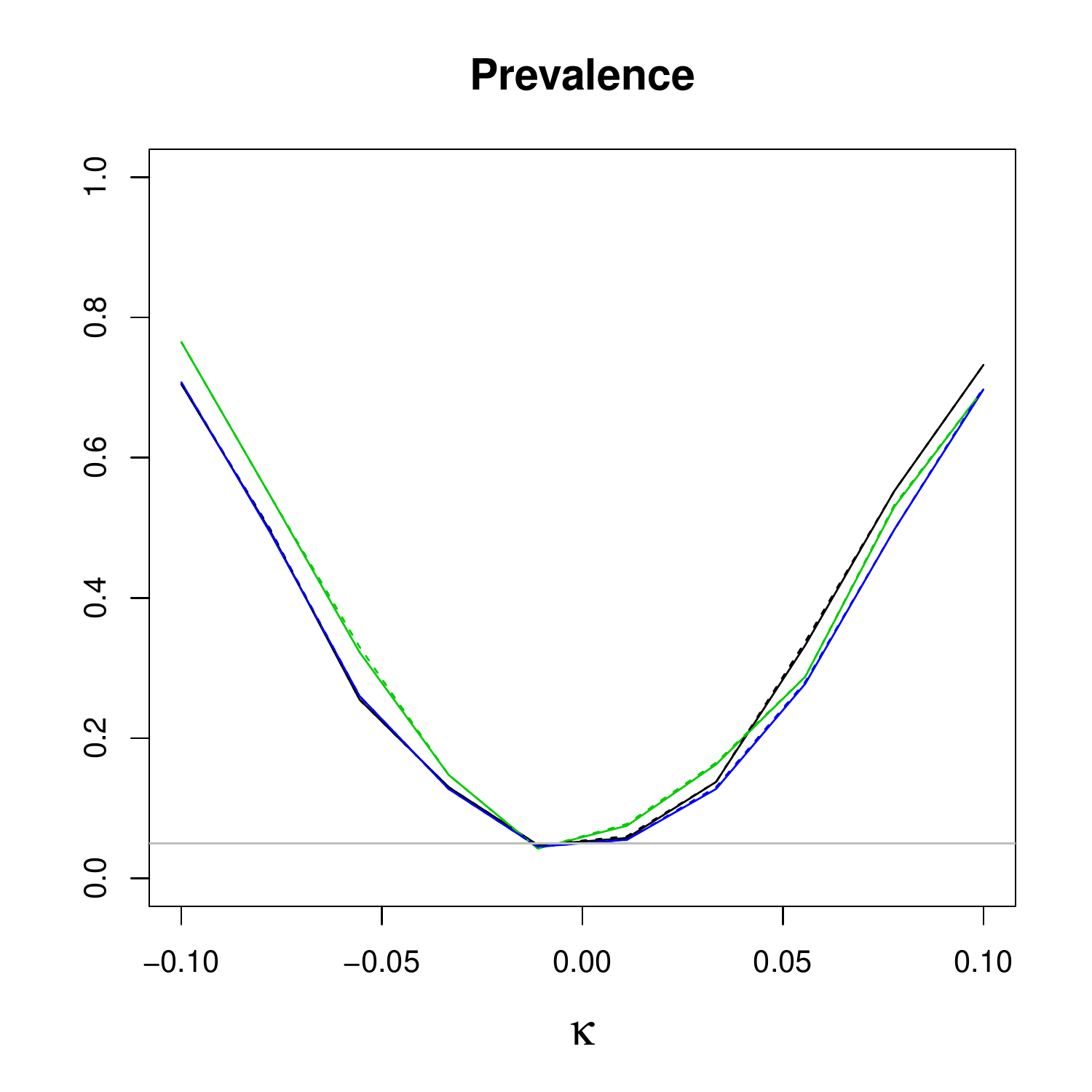}}%
\hspace{.01\textwidth}%
 \caption{Estimated power functions for constant (black), crossing (green), and deviating (blue) hazards, based on 250 subjects with a replication number of 400. The dashed lines show test statistics derived from existing methods in the literature, that are tailor-made for the particular scenario. The confidence level is shown by the gray lines.}
 \label{fig:power functions censoring}
\end{figure}

In each scenario, we rejected $\textbf{H}^{X}_0$ at the 5\% confidence level. Thus, we obtained $m$ Bernoulli trials in which the success probability is the power function evaluated at $\kappa$. The estimated power functions, i.e.\ the estimates of the Bernoulli probabilities, are displayed in figure \ref{fig:power functions censoring} (solid lines). The power functions are not affected by the structure of the underlying hazards, as desired:\ our tests are only defined at $t_0$, and the particular parameterization of the hazard has minor impact on the power function.

The dashed lines in figure \ref{fig:power functions censoring} show power functions of alternative nonparametric test statistics that are already found in the literature, tailor-made for the scenario of interest. In particular, for the survival at $t$, we obtained a test statistic using Greenwood's variance formula (and a cloglog transformation in the Supplementary Material) \cite{klein2007analyzing}. For the restricted mean survival at $t$, we used the statistic suggested in \cite{tian2017efficiency}. For the cumulative incidence at $t$, we used the statistic suggested in \cite{zhang2008summarizing}. For the mean frequency function we used the estimators derived in \cite{ghosh2000nonparametric}, and in the prevalence example we used the variance formula in \cite[p. ~ 295]{Andersen}, as implemented in the \texttt{etm} package in \texttt{R}. 
Our generic strategy gave similar power compared to the conventional methods 
for each particular scenario. 


\subsection{Comparisons with the log-rank test}
\begin{figure}
\centering
\includegraphics[width=1.1\textwidth]{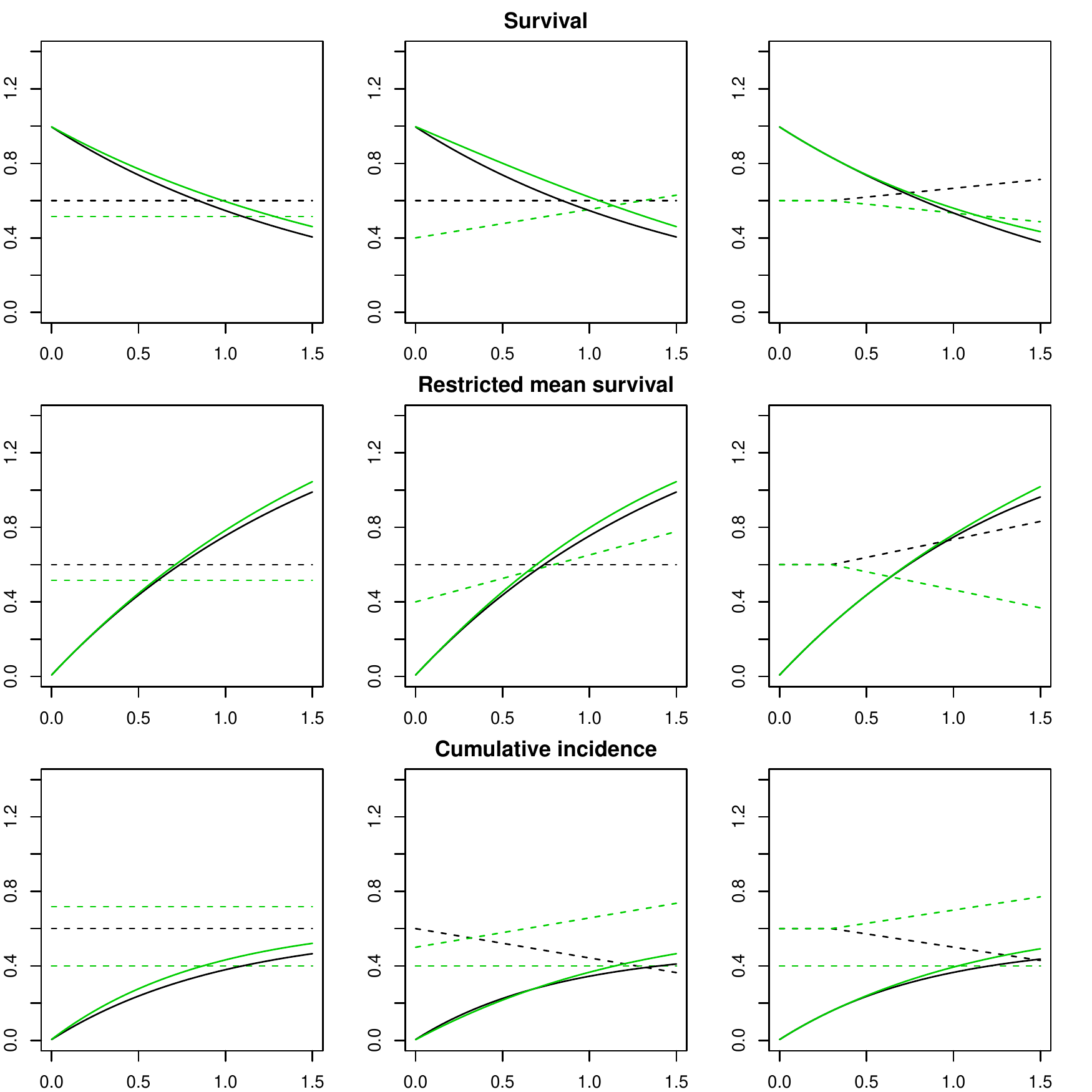}
 \caption{Simulation scenarios in which the true parameter difference was fixed to $\kappa=-0.05$ at $t_0 = 1.5$, i.e. $X^1_{1.5} -X^2_{1.5} = -0.05$. The upper row shows survival, the middle row shows restricted mean survival, and lower row shows cumulative incidences. The hazards are displayed as dotted lines; constant in the left column, linearly crossing in the middle column, and deviating in the right column. The $X^1$ parameters/hazards are black, and the $X^2$ parameters/hazards are green. See Table \ref{table: power comparison} for a power comparison of our tests and the rank tests for the scenarios that are displayed. The cumulative incidence panels: The cause specific hazards for the competing event are held constant equal to $0.4$ at all times. We optimize the cause specific hazards for the event of interest so that $X_{1.5}^1 - X_{1.5}^2 = -0.05$ under the restrictions that they are constant (left), linearly crossing (middle), and equal before deviating (right). }
 \label{fig:simulation scenarios}
\end{figure}

We have argued that our tests are fundamentally different from the rank tests, as the null hypotheses are different. Nevertheless, since rank tests are widely used in practice, also when the primary interest seems to be other hypothesis than in \eqref{eq: H0 log-rank}, we aimed to compare the power of our test statistics with the log-rank test under different scenarios. In table \ref{table: power comparison}, we compared the power of our test statistic and the rank test, using the scenarios in figure \ref{fig:simulation scenarios}. 
In the first column, the proportional hazards assumption is satisfied (constant), and therefore the power of the log-rank test is expected to be optimal (assuming no competing risks). Our tests of the survival function and the restricted mean survival function show only slightly reduced power compared to the log rank test. For the cumulative incidence function at a time $t_0$, our test is less powerful than the Gray test and the log-rank test of the cause specific hazards. However, the cause specific hazard test have type one error rate is not nominal, which we will return to in the end of this section.

The second column displays power of tests under scenarios with crossing hazards. For the survival function, it may seem surprising that the log-rank test got higher power than our test, despite the crossing hazards. However, in this particular scenario the hazards are crossing close to the end of the study (dashed lines in Figure \ref{fig:simulation scenarios}), and therefore the crossing has little impact on the power of the log-rank test. In contrast, the power of the log-rank test is considerably reduced in the scenarios where we study the restricted mean survival function and the cumulative incidence functions, in which the hazards are crossing at earlier points in time. 

The third column shows the power under hazards that are deviating. For the survival function, our test provides higher power. Intuitively, the log-rank test has less power in this scenario because the hazards are equal or close to equal during a substantial fraction of the follow-up time. For the restricted mean survival, however, the log-rank has more power. This is not surprising \cite{tian2017efficiency}, and it is due to the particular simulation scenario: Late events have relatively little impact on the restricted mean survival time, and in this scenario a major difference between the hazards was required to obtain $\kappa$. Since the log-rank test is sensitive to major differences in the hazards, it has more power in this scenario. For the cumulative incidence, in contrast, the power of the log-rank test is lower than the power of our test. 

The results in table \ref{table: power comparison} illustrate that power depends strongly on the hazard scenario for the log-rank test, but this dependence is not found for our tests. 

\begin{figure}
\centering
\includegraphics[width=1.1\textwidth]{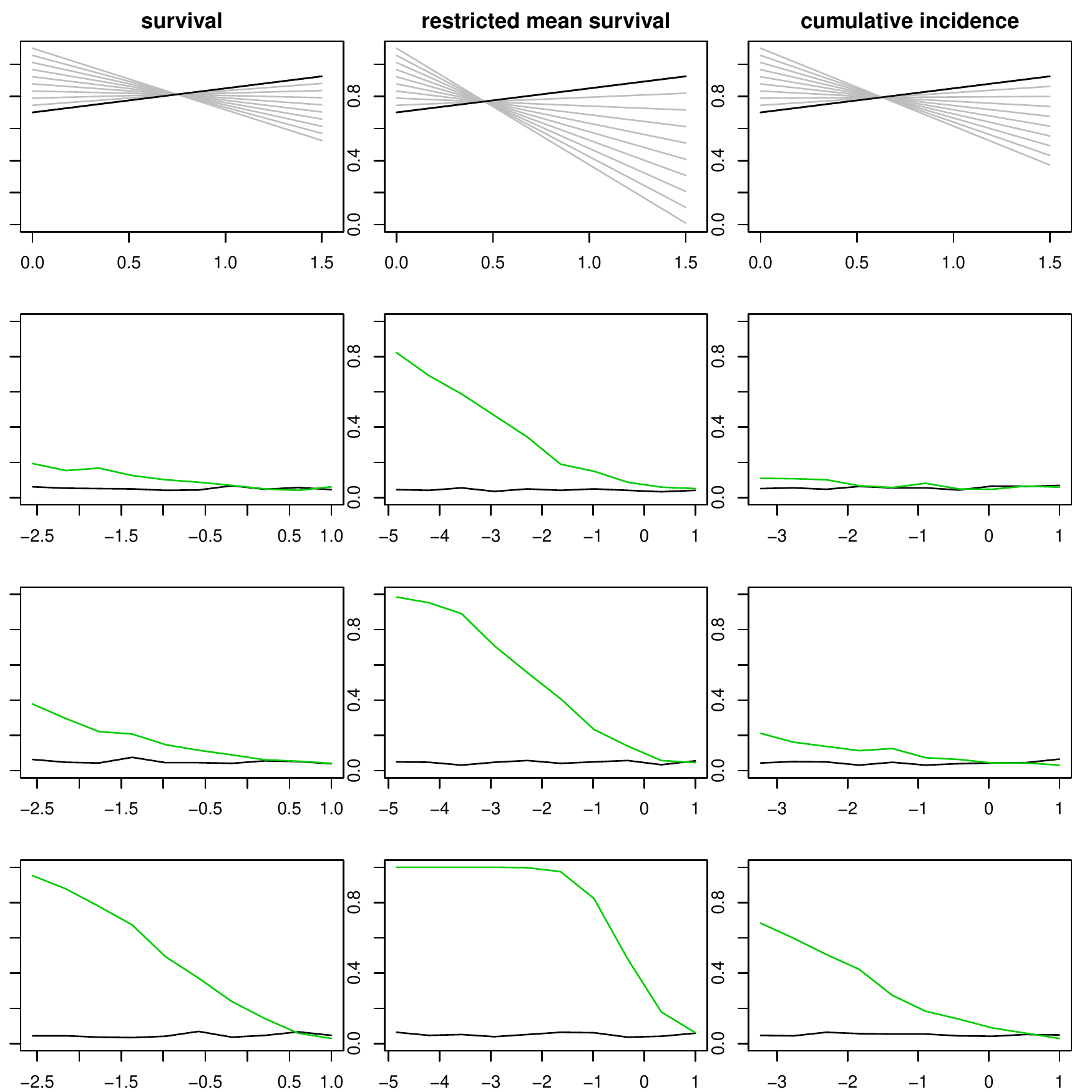}
 \caption{In the upper row, we display hazards functions in scenarios in which the hazard in group 1 is fixed (black line), and the hazards in group 2 varies (grey lines). The hazards are optimized such that the null hypothesis is true, i.e. $X^1_{t_0} = X^2_{t_0}$ for each combination of black/gray hazards at $t_0 = 1.5$. In the lower rows we show the estimated rejection rate as a function of the ratio of the hazard slopes (slope of gray/slope of black). This is done for sample sizes sample sizes 500 (row 2), 1000 (row 3), and 5000 (row 4) with a replication number of 500. The green curve shows the rejection rate of the log-rank test, while the black curve shows the rejection rate of our tests, which appear to be well calibrated along the 5\% significance level. If the sample size is large, the rank tests can falsely reject the null hypothesis even when the hazards are crossing. The cumulative incidence panels: We only show the cause-specific hazards for the event of interest (which we compare using the rank test). The cause-specific hazard for the competing event is equal to 0.4 in both groups. }
 \label{fig:simulation scenarios calibrated}
\end{figure}

To highlight the basic difference between the log-rank test and our tests, we have studied scenarios where $\textbf{H}^{X}_0$ in \eqref{eq: H0 prespecified t} is true (figure \ref{fig:simulation scenarios calibrated}). That is, at $t_0 = 1.5$ the difference $X_{t_0}^1 - X_{t_0}^2 = 0$, but for earlier times the equality does not hold. In these scenarios, the log-rank test got high rates of type 1 errors. Heuristically, this is expected because the hazards are different at most (if not all) times $t \in [0,t_0]$. Nevertheless, figure \ref{fig:simulation scenarios calibrated} confirms that the log-rank test does not have the correct type 1 error rate under null hypotheses as in \eqref{eq: H0 prespecified t}, and should not be used for such tests, even if the power sometimes is adequate (as in table \ref{table: power comparison}). 






\section{Example: Adjuvant chemotherapy in patients with colon cancer}
\begin{figure}
\centering
    \includegraphics[width=1.2\textwidth]{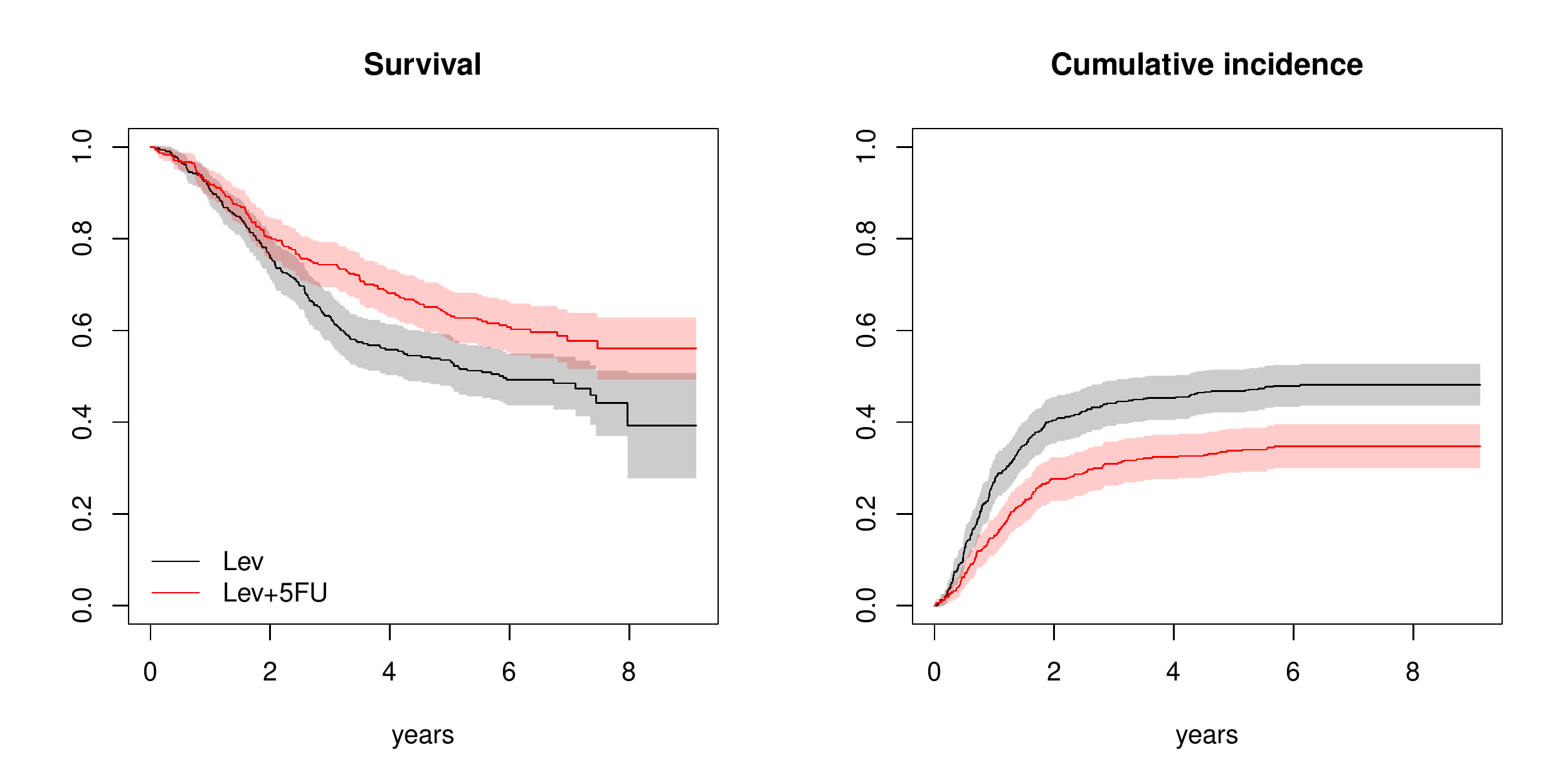}
 \caption{Survival curves (left) and the cumulative incidence of recurrence (right) along with 95\% pointwise confidence intervals (shaded) from the colon cancer trial are displayed.}
 \label{fig:survplot_colon}
\end{figure}

To illustrate how our tests can be applied in practice, we assessed the effectiveness of two adjuvant chemotherapy regimes in patients with stage III colon cancer, using data that are available to anyone \cite{moertel1995fluorouracil, survivalR}. The analysis is performed as a worked example in the supplementary data using the \texttt{R} package \texttt{transform.hazards}. After undergoing surgery, 929 patients were randomly assigned to observation only, levamisole (Lev) or levamisole plus fluorouracil (Lev+5FU) \cite{moertel1995fluorouracil}. We restricted our analysis to the 614 subjects who got Lev or Lev+5FU. All-cause mortality and the cumulative incidence of cancer recurrence was lower in subjects receiving (Lev+5FU), as displayed in Figure \ref{fig:survplot_colon}. 

We formally assessed the comparative effectivness of Lev and Lev+5FU after 1 and 5 years of follow-up, using the parameters from section \ref{sec: test statistic examples}. After 1 year, both the overall survival and the restricted mean survival were similar in the two treatment arms (Table \ref{table: 1 year}). However, the cumulative incidence of recurrence was reduced in the Lev+5FU group, and the number of subjects alive with recurrent disease were lower in the Lev+5FU group. Also, the mean time spent alive and without recurrence was longer in the Lev+5FU group (Table \ref{table: 1 year}, Restricted mean recurrence free survival). These results suggest that Lev+5FU has a beneficial effect on disease recurrence after 1 year of follow-up compared to Lev.  

After 5 years of follow-up, overall survival and restricted mean survival was improved in the Lev+5FU group (Table \ref{table: 5 year}). Furthermore, the cumulative incidence of recurrence was reduced, and the prevalence of patients with recurrence was lower in the Lev+5FU group. These results suggest that Lev+5FU improves overall mortality and reduces recurrence after 1 year compared to Lev.

In conclusion, our analysis indicates that treatment with Lev+5FU  improves several clinically relevant outcomes in patients with stage III colon cancer. We also emphasise that a conventional analysis using a proportional hazards model would not be ideal here, as the plots in Figure \ref{fig:survplot_colon} indicate violations of the proportional hazards assumptions. 


\section{Covariate adjustments}
\label{sec: covariate adjustments}
Our approach allows us to conduct tests conditional on baseline covariates, using the additive hazard model; by letting the cumulative hazard integrand in \eqref{eq:ode} be conditional on specific covariates, we can test for conditional differences between groups, assuming that the underlying hazards are additive. 


In more detail, we can test for differences between group 1 and 2 under the covariate level $Z=z_0$ by evaluating the cumulative hazards in each group at that level, yielding $A^{1,z_0}$ and $A^{2,z_0}$. Estimates $\hat A^{1,z_0}$ and $\hat A^{2,z_0}$ can be found using standard software. This allows us to estimate parameters with covariances using \eqref{eq:sde} and \eqref{eq:XVarEst}, and test the null hypothesis of no group difference within covariate level $z_0$ using the test statistic \eqref{eq: prespecified t statistic independence}, again assuming that the groups are independent.


\section{Discussion}

By expressing survival parameters as solutions of differential equations, we provide generic hypothesis tests for survival analysis. In contrast to the conventional approaches that are based on hazard functions \cite[Section 3.3]{aalen2008survival}, our null hypotheses are defined with respect to explicit parameters, defined at a time $t_0$. Our strategy also allows for covariate adjustment under additive hazard models. 

We have presented some examples of parameters, and our simulations suggest that the test statistics are well-behaved in a range of scenarios. Indeed, for common parameters such as the survival function, the restricted mean survival function and the cumulative incidence function, our tests obtain similar power to conventional tests that are tailor made for a particular parameter. Importantly, our examples do not comprise a comprehensive set of relevant survival parameters, and several other effect measures for event histories may be described on the differential equation form \eqref{eq:ode}, allowing for immediate implementation of hypothesis tests, for example using the \texttt{R} package \texttt{transforming.hazards} \cite{ryalen2017transforming,ryalen2018additive}. The fact that our derivations are generic and easy to implement for customized parameters, is a major advantage.


Our tests differ from the rank tests, as the rank tests are based on assessing the equality of the hazards during the entire follow-up. However, our strategy is intended to be different: We aimed to provide tests that apply to scenarios where the null hypothesis of the rank tests is not the primary interests. 


Restricting the primary parameter to a single time $t_0$ is sometimes considered to be a caveat. In particular, we ignore the time-dependent profile of the parameters before and after $t_0$. For some parameters, such as the survival function or the cumulative incidence function, this may be a valid objection in principle. However, even if our primary parameter is assessed at $t_0$, this parameter may account for the whole history of events until $t_0$. One example is the restricted mean survival, which considers the history of events until $t_0$. Indeed, the restricted mean survival has been suggested as an alternative effect measure to the hazard ratio, because it is easier to interpret causally, and it does not rely on the proportional hazards assumption \cite{trinquart2016comparison}. An empirical analysis of RCTs showed that tests of the restricted mean survival function yield results that are concordant with log-rank tests, under the conventional null hypothesis in \eqref{eq: H0 log-rank} \cite{trinquart2016comparison}, and similar results were found in an extensive simulation study \cite{tian2017efficiency}.

Furthermore, the time-dependent profile before $t_0$ is not our primary interest in many scenarios. In medicine, for example, we may be interested in comparing different treatment regimes, such as radiation and surgery for a particular cancer. Then, time to treatment failure is expected to differ in the shorter term due to the fundamental difference between the treatment regimes, but the study objective is to assess longer-term treatment differences \cite{ryalen2018causal}. Similarly, in studies of cancer screening, it is expected that more cancers are detected early in the screening arm, but the scientist's primary aim is often to assess long-term differences between screened and non-screened. In such scenarios, testing at a prespecified $t_0$ are more desirable than the null-hypothesis of the rank tests.  

Nevertheless, we must assure that cherry picking of $t_0$ is avoided. In practice, there will often be a natural value of $t_0$. For example, $t_0$ (or multiple $t_1, t_2, \cdots,t_k$) can be prespecified in the protocol of clinical trials. In cancer studies, a common outcome is e.g.\ five year survival. Alternatively, $t_0$ can be selected based on when a certain proportion is lost to censoring. Furthermore, using confidence bands, rather than pointwise confidence intervals and p-values, is an appealing alternative when considering multiple points in time. There exist methods to estimate confidence bands based on wild bootstrap for competing risks settings \cite{lin1997non}, which were recently extended to reversible multistate models allowing for illness-death scenarios with recovery \cite{bluhmki2018wild}. We aim to develop confidence bands for our estimators in future research. 

Finally, we are often interested in assessing causal parameters using observational data, under explicit assumptions that ensure no confounding and no selection bias. Such parameters may be estimated after weighting the original data \cite{ryalen2018additive,hernan2000marginal}. Indeed, weighted point estimators are consistent when our approach is used \cite{ryalen2018additive}, but we would also like to identify the asymptotic root $n$ residual distribution, allowing us to estimate covariance matrices that are appropriate for the weighted parameters. We consider this to be an important direction for future work. Currently, such covariance matrices can only be obtained from bootstrap samples.



\section{Software}
We have implemented a generic procedure for estimating parameters and covariance matrices in an \texttt{R} package, available for anyone to use on 
\texttt{github.com/palryalen/transform.hazards}. It allows for hypothesis testing at prespecified time points using \eqref{eq: H0 prespecified t}. 
Worked examples can be found the package vignette, or in the supplementary material here.

\section{Funding}
The authors were all supported by the research grant NFR239956/F20 - Analyzing clinical health registries: Improved software and mathematics of identifiability.

\bibliography{references}
\bibliographystyle{unsrt}

\clearpage

\begin{table}
\centering
\caption{Estimates, 95\% confidence intervals and p-values after 1 year}
\label{table: 1 year}
\begin{tabular}{c|c|c|c}
 & Lev & Lev+5FU &  p-value \\
\hline
Survival & 0.91 (0.87,0.94)  & 0.92 (0.89,0.95) & 0.62 \\
Restricted mean survival & 0.96 (0.95,0.98)  & 0.97 (0.95,0.98) & 0.86 \\
Cumulative incidence & 0.27 (0.22,0.32)  & 0.15 (0.11,0.19) & 0 \\
Prevalence & 0.19 (0.14,0.23)  & 0.09 (0.05,0.12) & 0 \\
Restricted mean recurrence free survival & 0.85 (0.82,0.88)  & 0.90 (0.87,0.92) & 0.01 \\
\end{tabular}
\end{table}

\begin{table}
\centering
\caption{Estimates, 95\% confidence intervals and p-values after 5 years}
\label{table: 5 year}
\begin{tabular}{c|c|c|c}
 & Lev & Lev+5FU & p-value \\
\hline
Survival & 0.54 (0.48,0.59)  & 0.63 (0.58,0.69) & 0.01 \\
Restricted mean survival & 3.62 (3.44,3.81)  & 3.97 (3.79,4.15) & 0.01 \\
Cumulative incidence & 0.47 (0.42,0.51)  & 0.34 (0.29,0.39) & 0 \\
Prevalence & 0.07 (0.05,0.10)  & 0.03 (0.02,0.05) & 0.02 \\
Restricted mean recurrence free survival & 2.29 (2.07,2.51)  & 2.95 (2.73,3.18) & 0 \\
\end{tabular}
\end{table}

\begin{table}
\centering
\caption{Power comparisons of our tests and rank tests (our/rank) for the scenarios displayed in figure \ref{fig:simulation scenarios}, comparing two groups of 1500 individuals (based on 400 replications). In the lower row, we also display the power of Gray's test for competing risks (our/rank/Gray). The power of the rank tests is sensitive to the shape of the underlying hazards, while the power of our tests vary little across the scenarios. In particular, the power of the rank tests is very sensitive to the rate of change of the hazards when they are crossing or deviating; see also the third column of figure \ref{fig:simulation scenarios}. }
\label{table: power comparison}

\begin{tabular}{c}
\end{tabular}
    \begin{tabular}{c|c|c|c}
    Parameter $\backslash$ Hazard & Constant & Crossing & Deviating  \\
    \hline
    Survival & 0.81/0.88& 0.79/0.96& 0.83/0.70\\  
    Restricted mean survival & 0.77/0.87& 0.78/0.21& 0.8/1\\  
    Cumulative incidence & 0.85/0.94/0.88& 0.86/0.80/0.70& 0.86/0.83/0.76\\  
    \end{tabular}
\end{table}

\clearpage

\appendix

\end{document}